\documentclass[twocolumn,twoside,slac_two]{revtex4}
\usepackage{graphicx}
\usepackage{fancyhdr}
\begin{document}

\title{MAXI monitoring of blazars and blackhole binaries}

%

\author{Juri Sugimoto}
\affiliation{MAXI team, Institute of Physical and Chemical Research (RIKEN), 2-1 Hirosawa, Wako, Saitama 351-0198, Japan}
\author{Hitoshi Negoro}
\affiliation{Department of Physics, Nihon University, 1-8-14, Kanda-Surugadai, Chiyoda-ku, Tokyo 101-8308}
\author{Satoshi Nakahira}
\affiliation{ISS Science Project Office, Institute of Space and Astronautical Science (ISAS),
 Japan Aerospace Exploration Agency (JAXA), 2-1-1 Sengen, Tsukuba, Ibaraki 305-8505, Japan}
\author{Yoshihiro Ueda}
\affiliation{Department of Astronomy, Kyoto University, Kitashirakawa-Oiwake-cho, Sakyo-ku, Kyoto 606-8502, Japan}
\author{Naoki Isobe}
\affiliation{Institute of Space and Astronautical Science (ISAS), Japan Aerospace Exploration Agency (JAXA), 
3-1-1 Yoshinodai, Chuo-ku, Sagamihara, Kanagawa 252-5210, Japan}
\author{Mutsumi Sugizaki}
\affiliation{MAXI team, Institute of Physical and Chemical Research (RIKEN), 2-1 Hirosawa, Wako, Saitama 351-0198, Japan}
\author{Tatehiro Mihara}
\affiliation{MAXI team, Institute of Physical and Chemical Research (RIKEN), 2-1 Hirosawa, Wako, Saitama 351-0198, Japan}
\author{Masaru Matsuoka}
\affiliation{MAXI team, Institute of Physical and Chemical Research (RIKEN), 2-1 Hirosawa, Wako, Saitama 351-0198, Japan}

\begin{abstract}
Since August 2009, MAXI experiment on the ISS has been performing all-sky X-ray monitoring. 
With MAXI, we detected flaring activities of some blazers, including Mrk 421, Mrk 501, and 3C 273.
Recently, new X-ray flaring activities were detected from two blazers, MAXI J1930+093 = 2FGL J1931.1+0938 \cite{5943} and 2MAXI J0243-582 = BZB J0244-5819 \cite{6012}.
The MAXI monitoring also covers black hole binaries, including  Cyg X--1 and Cyg X--3 which emit GeV gamma-rays. 
Their gamma-ray emission was found to coincide with their X-ray state transitions.
We present light curves and outstanding events of these sources.
\end{abstract}

\maketitle

\thispagestyle{fancy}



\section{MAXI}
We use the observations of Monitor of All sky X-ray Image (MAXI) \cite{matsuoka}.
MAXI was launched in 2009 July and attached to the International Space Station (ISS).
The ISS with MAXI orbits the earth in 92 minutes, and MAXI scans the objects in the all sky once in an orbit. 
MAXI has already reported more than one hundred transients\footnote{\url{http://maxi.riken.jp/top/doc/maxi_atel.html}}.
The observed results are immediately released through the internet, promoting rapid follow-up observations with telescopes around the world.
MAXI has two kinds of X-ray cameras: 
the Gas Slit Camera (GSC: \cite{mihara}) covering the energy band of 2--20 keV.
and the Solid-state Slit Camera (SSC: \cite{tomida,tsunemi}) covering 0.7--7 keV.
We can downloaded one-day bin and 90 min bin archival data for making light curves from the MAXI home page\footnote{\url{http://maxi.riken.jp/}}.
The energy spectra, images and also light curves of both the GSC and the SSC can be processed by the MAXI on-demand data web page\footnote{\url{http://maxi.riken.jp/mxondem/}} \cite{nakahira}.

\section{Observation of blazars for 5 years}

\begin{figure*}[t]
\centering
\includegraphics[width=8cm]{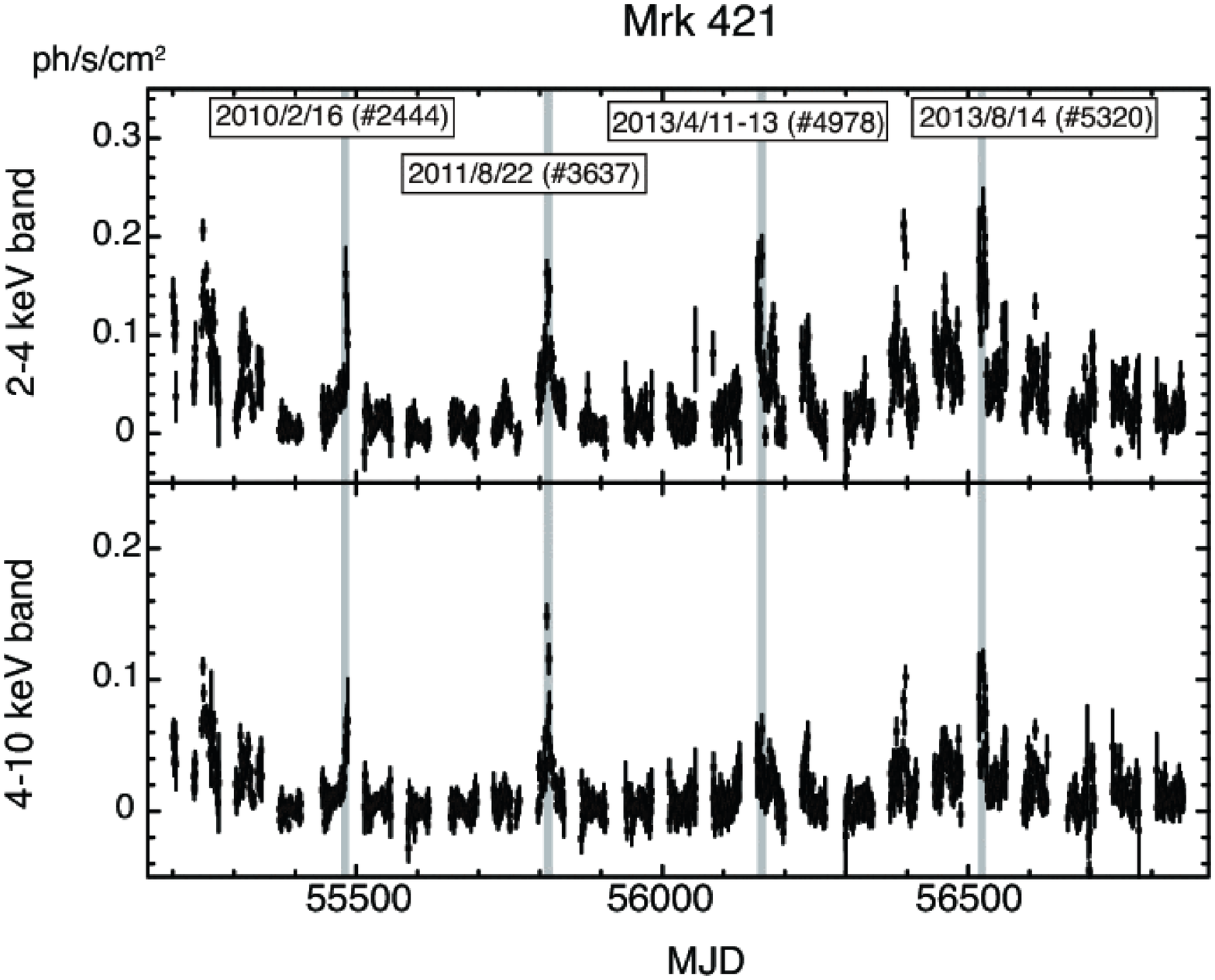}
\includegraphics[width=8cm]{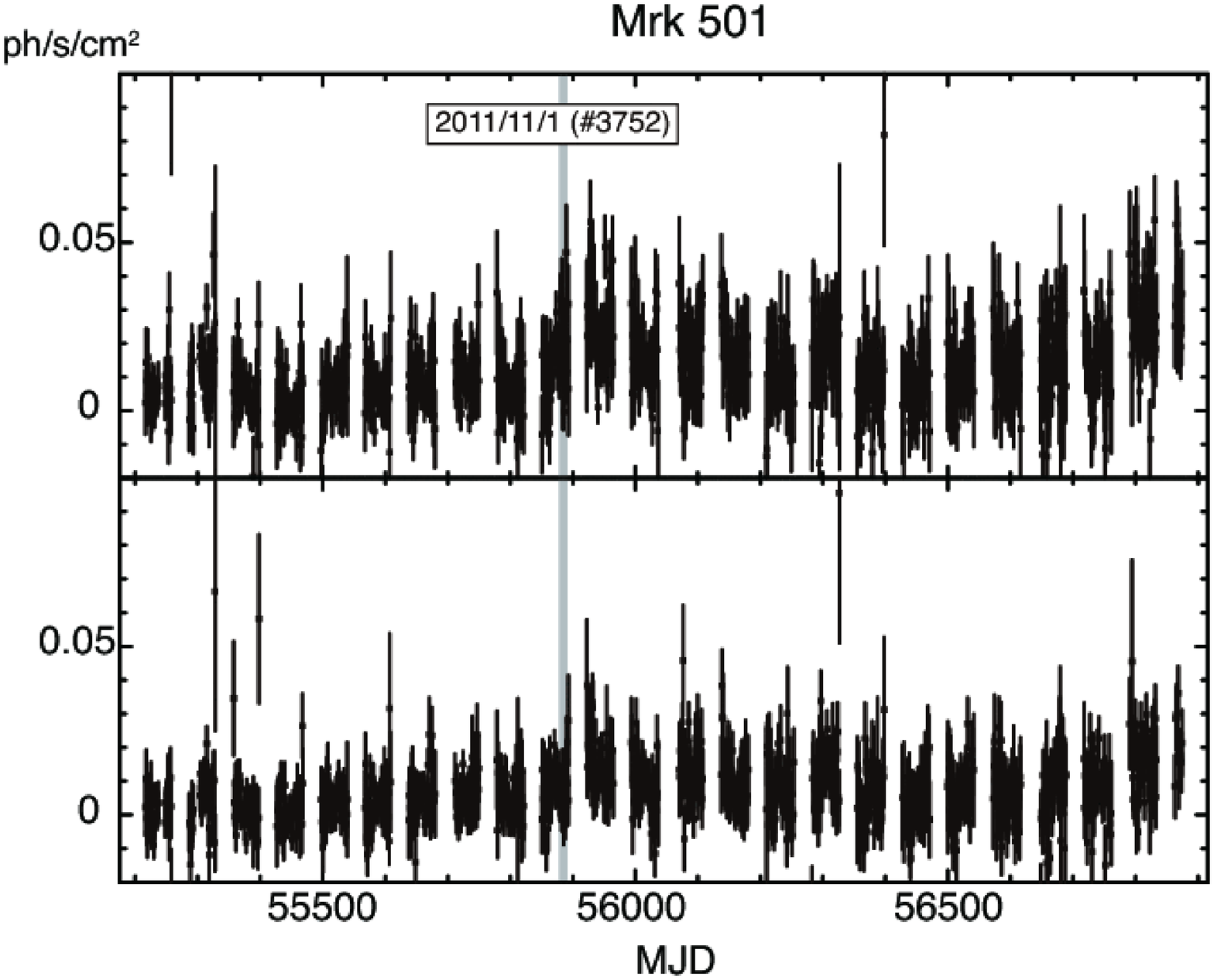}
\includegraphics[width=8cm]{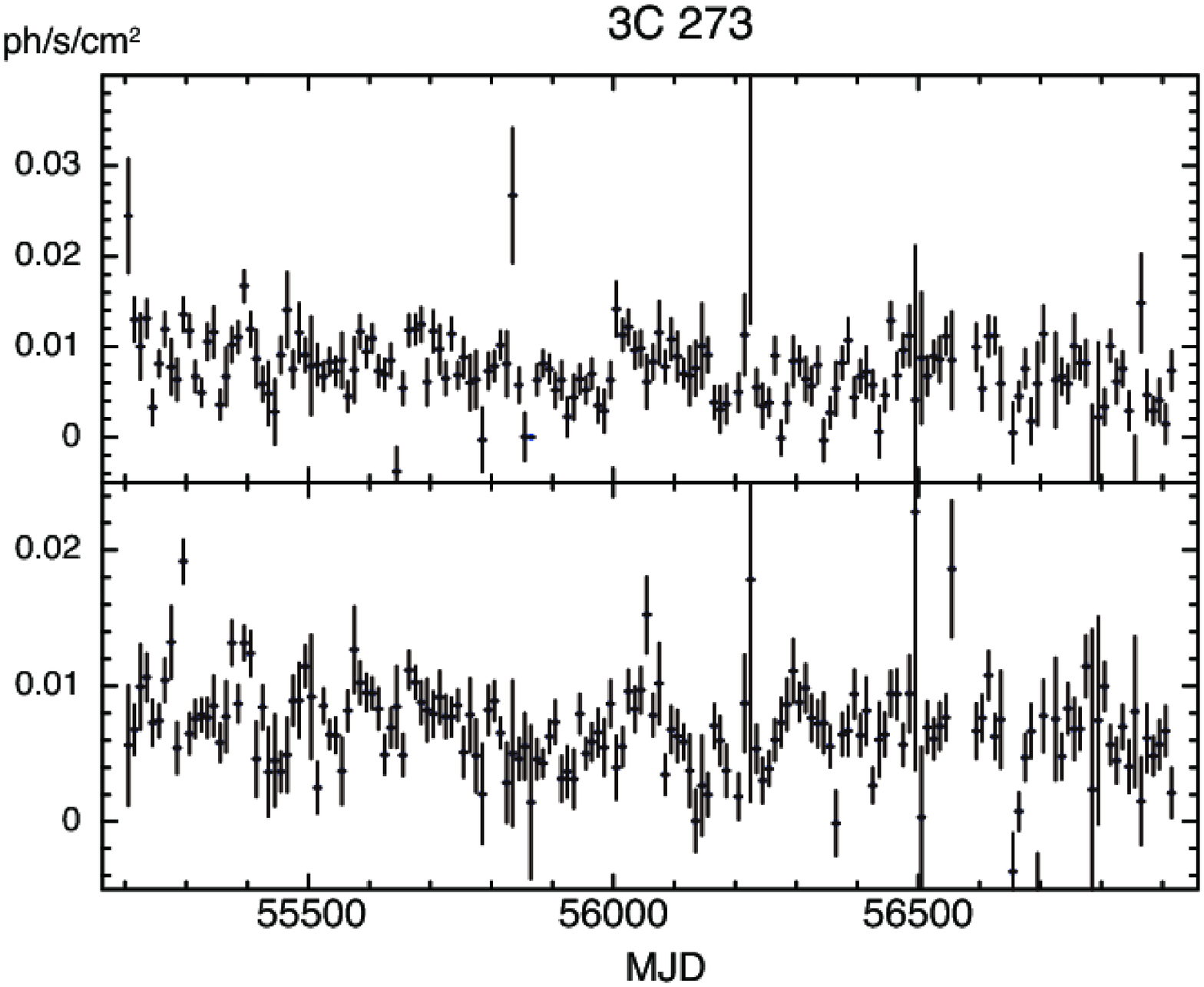}
\caption{MAXI/GSC light curves of blazars, Mrk 421, Mrk 501 and 3C 273.
Binning is 1 day for Mrk 421 and Mrk 501 and 10 days for 3C 273.
The dates in the figure are MAXI notifications to Atel.} 
\label{mrk421}
\end{figure*}

MAXI is monitoring 21 BL Lacs and 3 quasars. 
We show the light curves for 5 years by MAXI/GSC of Mrk 421, Mrk 501 and 3C 273 in Figure \ref{mrk421}. 
We reported their X-ray flares and brightenings to the Astronomer's Telegram (Atel, showed as grey lines in Figure \ref{mrk421}).

\section{New detection of two blazars}

\subsection{2FGL J1931.1+0938}
\begin{figure}[t]
\centering
\includegraphics[width=8cm]{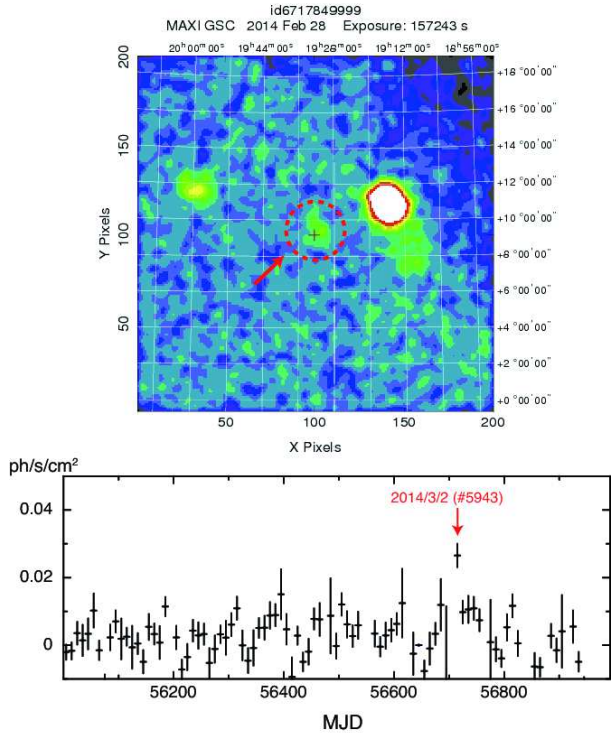}
\caption{The upper panel is a trigger image of 2FGL J1931.1+0938 by MAXI/GSC in 4--10 keV band, shown with a red arrow.
The lower panel shows the light curve in 2--10 keV energy band.
1 bin is 10 days.} 
\label{1930}
\end{figure}

At 05:31:55 UT on March 2, 2014, MAXI nova alert system detected a new faint X--ray source which was $\sim$10 mCrab in Figure \ref{1930}.
We named it MAXI J1930+093 and reported to the Atel$\#$5943 \cite{5943}.
After that, Swift/XRT observed the error region of MAXI. 
Swift found an X-ray source which was the same intensity as MAXI observation, and identified it to the BL Lac object 2FGL J1931.1+0938.

\subsection{BZB J0244--5819}
\begin{figure}[t]
\centering
\includegraphics[width=8cm]{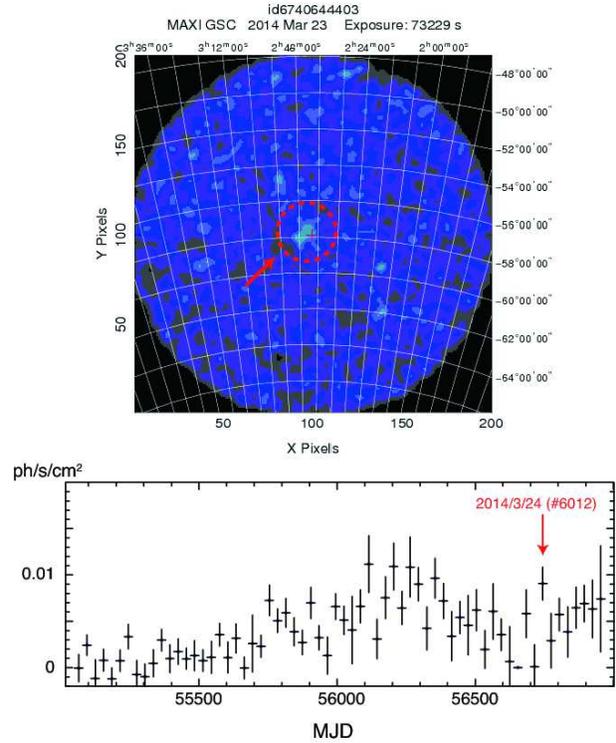}
\caption{The upper panel is a trigger image of BZB J0244-5819 by MAXI/GSC in 4--10 keV band, shown with a red arrow.
The lower panel shows the light curve in 2--10 keV energy band.
1 bin is 30 days.} 
\label{bzb}
\end{figure}

At 19:24:10 UT on March 24, 2014, MAXI nova alert system detected transient object which was 6.6 mCrab in Figure \ref{bzb}. 
It had been identified as 2MAXI J0243--582 in the MAXI/GSC 37-Month catalog \cite{hiroi}. 
We proposed a Swift ToO observation with 4-point tiling to cover the MAXI error circle with the Swift XRT. 
In the Swift XRT image, we find a bright point source at (RA, Dec)= (2h 44m 40.10s, $-$58d 19m 54.8s) with an estimated error of 2.3 arcseconds radius (90$\%$ c.l.). 
This position lies 1.54 arc-seconds from the NED position of the BL Lac object BZB J0244--5819. 
We therefore suggest that the trigger source is an X-ray flare of BZB J0244--5819 (=MAXI J0243--582) (ATel$\#$6012 \cite{6012}).


\section{Black hole binary Cyg X-1}
Cygnus X-1 (Cyg X--1) is one of the most famous high mass X-ray binaries (HMXBs), 
and is composed of a black hole (BH) and a massive giant companion star.
X--ray from Cyg X--1 is highly variable,the binary period is 5.6 d, and the distance is 1.86$^{+0.12}_{-0.11}$ kpc \cite{orosz}.
The galactic BH binaries have two spectral states, a low/hard state that is dominated  by a power-law spectrum, corresponding to the radiatively inefficient accretion flow (RIAF), 
and a high/soft state that is dominated by a thermal emission from the standard optically thick accretion disk \cite{remillard2006,done}.

\subsection{Light curve}
MAXI obtained a long-term light curve for more than 5 years of Cyg X--1 \cite{sugimoto}.
Cyg X--1 had been in the low/hard state until June 2010, 
and after that it stayed in the high/soft state for about ten months \cite{Negoro10}.

\begin{figure*}[htpb]
  \begin{center}
   \includegraphics[width=12.5cm]{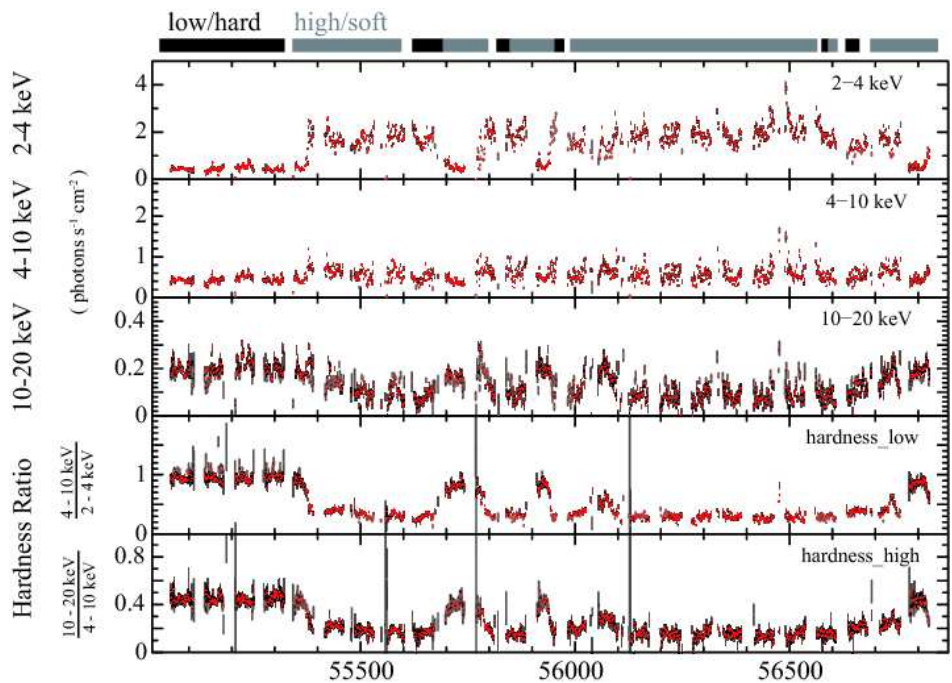}
  \end{center}
  \caption{One-day bin light curves and hardness ratios of Cyg X--1 obtained with MAXI/GSC.
  From top to bottom panel, light curves in the energy bands: 2--4 keV, 4--10 keV and 10--20 keV,
  and the hardness ratios of $I$(4--10 keV)/$I$(2--4 keV) and $I$(10--20 keV)/$I$(4--10 keV) are shown.
  The black and grey regions show the terms in the low/hard state, and in the high/soft state, respectively.}
\label{fig:cygLC}
\end{figure*}

Figure \ref{fig:cygLC} shows light curves with one-day bin of Cyg X--1 obtained with GSC from 15 August 2009 (55058 MJD) to 9 November 2014 (56970 MJD),
in three energy bands (2--4 keV, 4--10 keV and 10--20 keV).
The two kinds of hardness ratios, $I$(4--10 keV)/$I$(2--4 keV) and $I$(10--20 keV)/$I$(4--10 keV), are also plotted.
The state of Cyg X--1 can be recognized by the values of the hardness ratios.

The low/hard state continued for about ten months since the start of the MAXI observation.
A transition to the high/soft state occurred around 55378 MJD and then continued for another ten-month.
After several state transitions, it has stayed in the high/soft state since 56107 MJD.

\subsection{Hardness-intensity diagram}

\begin{figure}[htbp]
  \begin{center}
   \includegraphics[width=6cm]{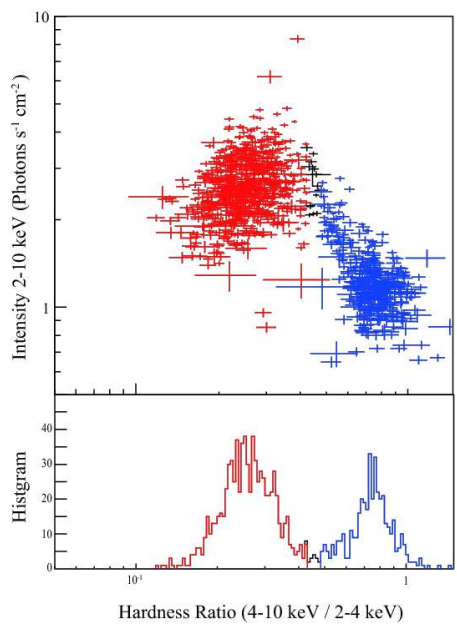}
  \end{center}
  \caption{Upper: the hardness-intensity diagram of Cyg X--1. 
  Lower: the histogram of the hardness ratio distribution.
  The hardness ratio is calculated by dividing 4--10 keV band count rate by 2--4 keV one.
  The blue data points are in the low/hard state and the red ones are in the high/soft state.
  We treat the black ones to be during the transition.}
\label{fig:cygHDI}
\end{figure}

The upper panel of Figure \ref{fig:cygHDI} shows a hardness-intensity diagram.
The vertical axis shows count rates in the 2--10 keV band, 
and the horizontal axis indicates the hardness ratios of the count rates in the 4--10 keV band to those in the 2--4 keV band.
The lower panel in Figure \ref{fig:cygHDI} shows a histogram of the number of data points in certain bins of the hardness ratio.
We can see clear two peaks, which correspond to the high/soft state and the low/hard state.
To separate the period into those two states, we fit the histogram with two gaussian functions,
 and determined the mean values and standard deviations of the gaussian functions.
Then we defined the state of each data point, by checking whether the hardness ratio of the data point is in $\pm$ 3$\sigma$ of the distributions.
Blue and red data points in Figure \ref{fig:cygHDI} are thus defined the low/hard state and the high/soft state, respectively.
Black points are between then and considered as the transition.
The determined terms of the states are summarized in Table \ref{tb:trans}.

\begin{table}[htbp]
\small
\caption{The terms of spectrum states}
\begin{center}
\begin{tabular}[b]{c c c}
\hline
spectrum state & start MJD & end MJD \\ 
\hline
hard & 55058 & 55376 \\ 
soft & 55378 & 55673 \\ 
hard & 55680 & 55788 \\
soft & 55789 & 55887 \\ 
hard & 55912 & 55941 \\
soft & 55943 & 56068 \\
hard & 56069 & 56076 \\
soft & 56078 & 56733 \\
hard & 56735 & 56741 \\
soft & 56742 & 56757 \\ 
hard & 56781 & 56824 \\
soft & 56854 & $\sim$ \\
\hline
\end{tabular}
\end{center}
\label{tb:trans}
\end{table}

\bigskip 
\begin{acknowledgments}
This work was supported by RIKEN Junior Research Associate Program.
\end{acknowledgments}

\bigskip 

\begin{thebibliography}{99}   
  \bibitem{done} 
  Done, C., Gierli{\'n}ski, M., \& Kubota, A.\ 2007, A$\&$A Rev., 15, 1 
  \bibitem{hiroi} 
  Hiroi, K., Ueda, Y., Hayashida, M., et al. 2013, ApJS, 207, 36 
  \bibitem{matsuoka} 
  Matsuoka, M., Kawasaki, K., Ueno, S., et al.\ 2009, PASJ, 61, 999
  \bibitem{mihara} 
  Mihara, T., Nakajima, M., Sugizaki, M., et al.\ 2011, PASJ, 63, 623
  \bibitem{6012} 
  Nakahira, S., Ueno, S., Tomida, H., et al.\ 2014, The Astronomer's Telegram, 6012, 1 
  \bibitem{nakahira} Nakahira, S., et al., 2013, Journal of Space Science Informatics Japan, 2, 29 
  \bibitem{5943}
   Negoro, H., Serino, M., Nakahira, S., et al.\ 2014, The Astronomer's Telegram, 5943, 1 
  \bibitem{Negoro10} 
  Negoro, H., Kawai, N., Kawasaki, Y.~U.~K., et al. 2010, The Astronomer's Telegram, 2711, 1 
  \bibitem{orosz} 
  Orosz, J.~A., McClintock, J.~E., Aufdenberg, J.~P., et al.\ 2011, ApJ, 742, 84
  \bibitem{remillard2006} 
  Remillard, R.~A., \& McClintock, J.~E.\ 2006, ARA$\&$A, 44, 49 
  \bibitem{sugimoto} 
  Sugimoto, J., Mihara, T., Matsuoka, M., et al.\ 2014, Suzaku-MAXI 2014: Expanding the Frontiers of the X-ray Universe, 222
  \bibitem{tomida} 
  Tomida, H., Tsunemi, H., Kimura, M., et al.\ 2011, PASJ, 63, 397 
  \bibitem{tsunemi} 
  Tsunemi, H., Tomida, H., Katayama, H., et al.\ 2010, PASJ, 62, 1371
\end{thebibliography}

\end{document}